# High sensitivity measurement of ULF, VLF and LF fields with Rydberg-atom sensor

MINGWEI LEI,[1] MENG SHI[1,*]

[1]*Key Laboratory of Space Utilization, Technology and Engineering Center for Space Utilization, Chinese Academy of Sciences, Beijing 100094, China*
*\* shimeng@csu.ac.cn*

**Abstract:** Fields with frequencies below megahertz are challenging for Rydberg-atom-based measurements, due to the low-frequency electric field screening effect that is caused by the alkali-metal atoms adsorbed on the inner surface of the container. In this paper, we investigate on electric fields measurements in the ULF, VLF and LF bands in a Cs vapor cell with built-in parallel electrodes. With optimization of the applied DC field, we achieve high-sensitive detection of the electric field at frequencies of 1kHz, 10kHz and 100kHz based on Rydberg-atom sensor, with the minimum electric field strength down to 18.0μV/cm, 6.9μV/cm and 3.0μV/cm, respectively. The corresponding sensitivity is 5.7 μV/cm/√Hz, 2.2μV/cm/√Hz and 0.95μV/cm/√Hz for ULF, VLF and LF fields, which is better than 1-cm dipole antenna. Besides, the linear dynamic range of Rydberg-atom sensor is over 50 dB. This work presents the potential to enable more applications that utilize atomic sensing technology in ULF, VLF and LF fields.

## 1. Introduction

Rydberg atoms with high polarizability and electric dipole moments, are sensitive to external electric field [1]. Significant progresses on Rydberg-atom sensor have been made to achieve an ultra-high sensitivity down to 55nV/cm/√Hz [2] and expand the operating frequency ranged from DC to THz [3-6]. Based on Autler Townes splitting or Stark shift of electromagnetically induced transparency (EIT) spectroscopy [7,8], Rydberg-atom sensor have explored the strength, frequency, phase, polarization, arrival angle and subwavelength imaging of electric fields [9-11].

Most experiments are focused on the measurement of microwave with frequency ranged from 300MHz to 300GHz. Recently, the study on measuring field with frequency below gigahertz based on the Stark effect of Rydberg atoms has started. The highest sensitivities of MHz fields are demonstrated as 0.96μV/cm/√Hz by heterodyne techniques [12-14]. The sensitivity of Rydberg-atom sensor for the electric fields with frequency below kilohertz is estimated to 3.4 μV/cm/√Hz by spectral noise [15], and a sensitivity of 67.9 μV/cm/√Hz is achieved for 100Hz field [16], which exceeds that of a conventional antenna of the same size.

However, high-sensitive measurement of the kHz fields based on Rydberg-atom sensor is less studied before. The fields in the Ultra Low Frequency (ULF, 0.3kHz-3kHz), Very Low Frequency (VLF, 3kHz-30kHz), and Low Frequency (LF, 30kHz-300kHz) bands have significant applications in fields of remote timing, auxiliary navigation as well as radio communication, due to its long propagation distances and strong penetrability of these fields. The conventional receiver for low-frequency electric field usually needs a large antenna with length in meters to improve its sensitivity, which limits its further application. In contrast, Rydberg-atom sensors have a compact size in centimeters. Therefore, it makes sense to develop a high-sensitivity Rydberg-atom sensors to measure kHz electric fields.

Because of the shielding effect of the alkali-metal-atom container on low-frequency electric fields, the fields oscillating at frequencies below megahertz can hardly be detected. By adding parallel electrodes located inside an atomic vapor cell, it is possible to conduct measurements of the electric field at kHz frequency in this study. By optimizing an auxiliary DC field, we successfully detect electric fields with strengths down to 18.0 μV/cm, 6.9 μV/cm, and 3.0 μV/cm for the 1kHz, 10kHz and 100kHz signal fields, respectively. Notably, our Rydberg-atom sensor achieves the corresponding sensitivities of 5.7 μV/cm/√Hz, 2.2 μV/cm/√Hz and 0.95 μV/cm/√Hz, and exhibits an impressive linear dynamic range of over 50dB at kHz frequencies. This advancement shows the potential for enabling a broadened applications in the ULF, VLF, and LF electric fields with Rydberg-atom sensor.

## 2. Experiment setup

Figure 1(a) is the relevant energy levels of Cs atom, where $6S_{1/2}$ is the ground state, $6P_{3/2}$ are the low-lying excited state, and $63D_{5/2}$ is the Rydberg state. Due to the presence of the external electric fields, the magnetic substates of Rydberg state would be splitting to $M_J$ = 1/2, 3/2, 5/2 energy levels, due to the Stark shift. To avoid the screening effect of alkali metal atoms adsorbed on the inner wall of the glass container, we placed parallel electrode plates with dimensions of 80mm(L)*26mm(W)*1mm(H) and a spacing of 18mm inside a Cs vapor cell with a diameter of 40mm and a length of 100mm, shown in Fig. 1(b). The experimental schematic diagram is illustrated in Fig. 1(c). Before entering the cell, the 852-nm laser was split to two beams as a probe laser and a reference laser, in order to reduce the intensity fluctuations of lasers. The probe laser and reference laser are locked to the $6S_{1/2}$ (F=4) → $6P_{3/2}$(F'=4,5) transition line of Cs atoms to obtain the an electromagnetically-induced transparency (EIT) spectrum without Doppler background, while the coupling laser is scanned near the $6P_{3/2}$→$63D_{5/2}$ transition line to further excite the Cs atoms to the Rydberg state $63D_{5/2}$. Furthermore, the power of the probe laser and reference laser are equal as 50 μW, with a diameter of 600μm, while the power of the coupling light was 15 mW, with a diameter of 1mm. Besides, the frequency of the probe laser and coupling laser can be both locked to an ultra-stable cavity (PDH) to reduce the noise from laser frequency. Besides, kHz signal fields and an auxiliary DC field are provided by a signal generator, which are combined through a power combiner and then introduced to the parallel electrode plates inside the vapor cell. In order to generate kHz electric fields with strength below 1uV/cm, a key-press attenuator with range of 0-90dB is introduced with the kHz signal fields. Detected with a balanced photodetector, the EIT transmission spectra are sent to an oscilloscope for analysis.

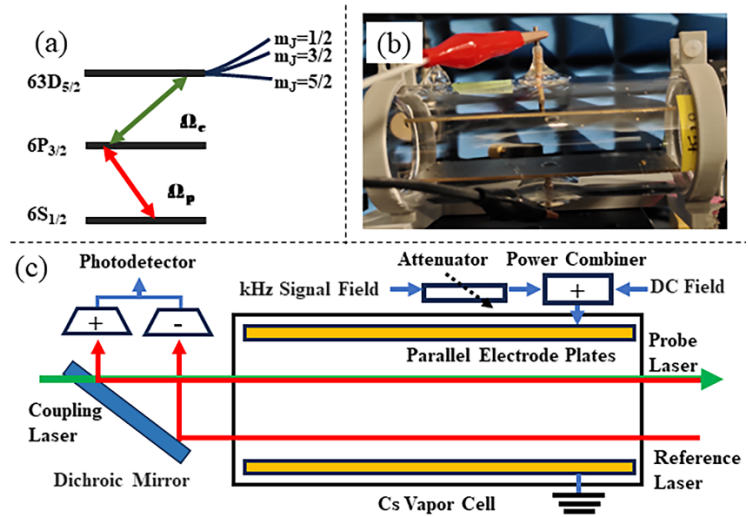

Fig. 1. (a) Relevant energy levels in cesium Rydberg EIT. (b) Photograph of Cs vapor cell with built-in parallel electrodes used in our experiments. (c) Schematic diagram of the experimental setup, including a vapor cell inside with parallel electrode plates, 852nm probe laser, 852nm reference laser, 509nm coupling laser, dichroic mirror, balanced photodetector, power combiner, DC field and kHz signal Field.

## 3. Results and discussion

Figure 2 shows the EIT spectra of Rydberg state $63D_{5/2}$ with or without the DC and 100kHz signal field into the ends of parallel electrode plates. Presented in Fig. 2, the green curve displays the field-free EIT spectrum by scanning the coupling laser, used for a reference EIT. When loaded with 100kHz signal field, distinct modulations emerge in the Rydberg EIT spectrum in the red curve of Fig. 2, as the transparency of

Cs atoms is modulated by electromagnetic fields. When loaded with a stronger DC field, the EIT spectrum exhibits splitting to three components (in the blue curve of Fig. 2), which reflects the breaking of degeneracy among fine structures of Rydberg state $63D_{5/2}$. This can be explained by the Stark shift of the magnetic substates as

$$\Delta f = -\frac{\alpha}{2}E^2, \tag{1}$$

herein $\alpha$ is the polarizability and $E$ is the strength of electric field. Calculated by Alkali Rydberg Calculator (ARC) [17], the polarizability of magnetic substates $M_J$ = 1/2, 3/2, 5/2 of Rydberg state $63D_{5/2}$ are different as -7312 MHz•cm$^2$/V$^2$, -5165 MHz•cm$^2$/V$^2$, 431 MHz•cm$^2$/V$^2$, respectively, which leads to their different Stark shift and then splitting to three spectral lines. When the 100kHz signal field and DC field are both loaded into the Rydberg-atom sensor, the modulations in the pink curve of Fig. 2 becomes more apparent and long-lasting than that only with 100kHz signal field. The two insets of fig. 2 clearly display that the modulations of Rydberg EIT spectra are in respond to the signal field, which are oscillates at the same frequency as 100 kHz. Furthermore, the response intensity to the 100kHz signal field is stronger when it is accompanied by a DC field compared to when the DC field is absent. When a kHz signal field is added to the electrodes together with the DC field, the total field can be expressed as

$$\varepsilon(t) = E_{DC} + E_{sig}e^{i(\omega_{sig}t + \varphi_{sig})}, \tag{2}$$

herein $E_{sig}$, $\omega_{sig}$, $\varphi_{sig}$ is, respectively, the strength, frequency and phase of the kHz signal field, $E_{DC}$ is the strength of the DC field. Under the weak electric field regime, the laser intensity after passing through Rydberg atoms can be written as

$$I(t) \propto E^2{}_{DC} + E^2{}_{sig} + 2E_{DC}E_{sig}\cos(\omega_{sig}t + \varphi_{sig}), \tag{3}$$

As a result, the modulation signals oscillate at the frequency $\omega_{sig}$, and is linear proportional to the strength $E_{sig}$ of the kHz signal field. According to Equation (3), the DC field acts as an amplifier of the kHz signal field with a coefficient of $2E_{DC}$. To achieve high sensitivity measurement of the kHz signal field, an auxiliary DC field can be introduced into Rydberg-atom sensor.

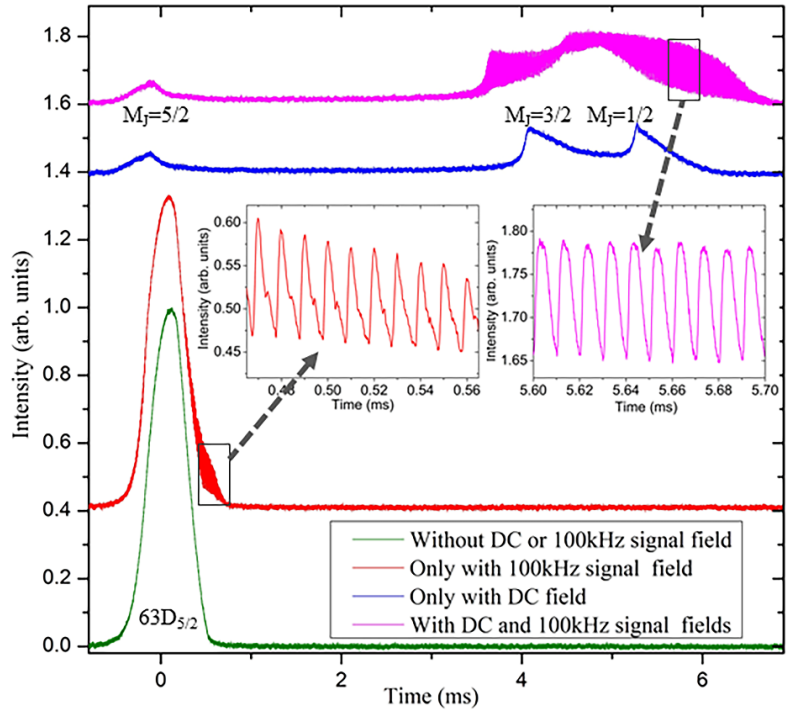

Fig. 2. EIT spectra for the Rydberg state $63D_{5/2}$ with or without the DC and 100kHz signal field. The left and right insets magnificently display the modulations of Rydberg EIT spectra in respond to 100kHz signal field without and with the DC and field, respectively.

To conduct electric field measurements by Rydberg-atom sensor, it is needed to calibrate the field strength, which can be calculated as

$$E = KV/Cd, \qquad (4)$$

where V is the amplitude of the voltage displayed on the signal generator, C represents an effective voltage coefficient, that $C = \sqrt{2}$ for the AC signal field and $C = 1$ for the DC field, d is the spacing between two electrode plates of 18mm in our vapor cell, K is a calibration coefficient. Using Eq. (1) and (4), the calibration coefficient K is obtained through linear fitting in a region of strong electric field, which is 0.458, 0.439, 0.443 and 1.185, respectively for 1kHz, 10kHz, 100kHz signal fields and DC field. The significant difference in the calibration coefficients between the kHz signal and the DC fields arises from the use of an attenuator with an insertion loss for the strength of the kHz signal fields. Further to achieve high sensitivity measurement of the kHz signal fields, we optimized the strength of the DC electric field to maximize the intensity of the kHz modulation signals received by the Rydberg atoms. Then, we conducted sensitivity tests for the kHz signal fields with the frequency of 1kHz, 10kHz and 100kHz, which represents measurements of ULF, VLF and LF bands, respectively. Using the spectrum analysis of the oscilloscope, the power of modulation signals oscillating at the same frequency of the kHz signal fields is measured under different strength of the signal field with a measurement integration time of 100ms. Presented in Fig. 3, the minimum detectable field at frequencies of 1kHz, 10kHz and 100kHz is down to 18.0μV/cm, 6.9μV/cm and 3.0μV/cm, and the linear dynamic range is 51dB, 63dB and 61dB, respectively.

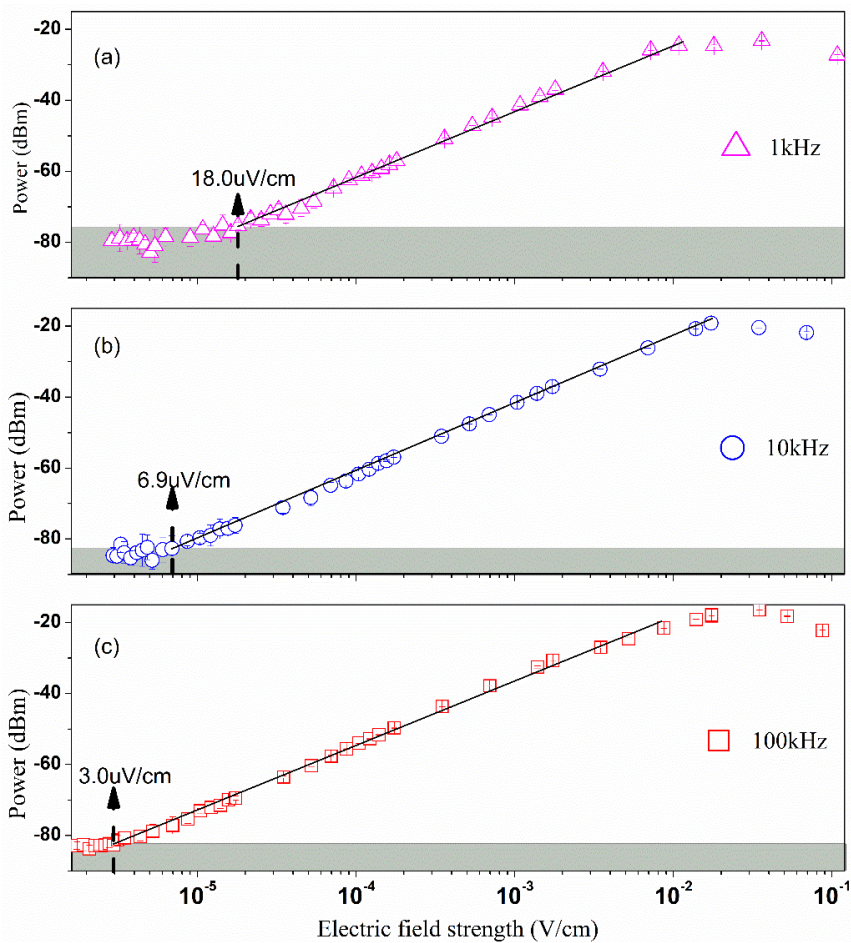

Figure 3. Strength measurement for the 1kHz (a), 10kHz (b) and 100kHz (c) signal fields using Rydberg-atom sensor. The data are taken by the spectrum analysis of the oscilloscope with an integration time of 100ms. Each data point is the average of six independent measurements, and the error bar is the standard error. The grey area represents the background noise.

The sensitivity of electric field measurements by Rydberg-atom sensor is estimated as
$$S = E_{min}/\sqrt{RBW}, \tag{5}$$
where $E_{min}$ is the minimum detectable field by the sensor, and RBW is the resolution bandwidth, which is 10 Hz in our measurements. The corresponding sensitivity of the Rydberg-atom sensor is 5.7μV/cm/√Hz, 2.2μV/cm/√Hz and 0.95μV/cm/√Hz for the 1kHz, 10kHz and 100kHz signal fields, respectively. For comparison, the sensitivity of the dipole antenna with a length of 1cm is calculated as a function of the frequency of electric field [4]. And the sensitivity of the reported Rydberg-atom sensors with frequency ranged from 100Hz to 200Mhz [12-16], is also attached in Fig. 4. For the kHz fields, the sensitivity of Rydberg-atom sensor is at the same level of ~1μV/cm/√Hz with the MHz fields. What is more, the sensitivity achieved with Rydberg-atom sensor is superior to the theoretical sensitivity limit of 1cm dipole antennas, and the advantage is more pronounced with lower frequency electric fields in ULF, VLF bands.

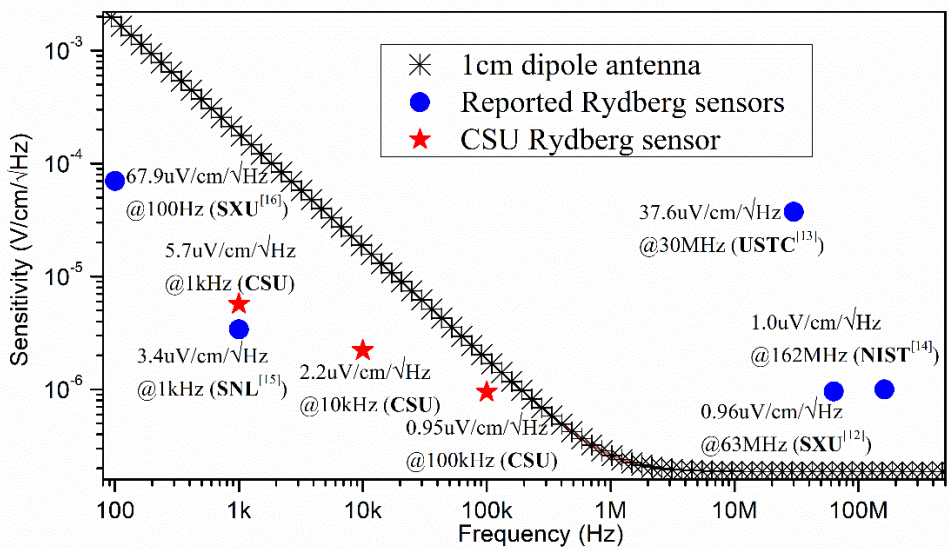

Fig. 4. Sensitivity as a function of the signal field frequency for the theoretical calculation of 1-cm dipole antenna, as well as for the reported Rydberg-atom sensors (blue circle) and our Rydberg-atom sensor (red star). Herein, CSU, SXU, USTC, SNL, NIST are, respectively, short for, Technology and Engineering Center for Space Utilization, Shanxi University, University of Science and Technology of China, Sandia National Laboratories, National Institute of Standards and Technology.

## 4. Conclusion

We have conducted an in-depth study on the application of Rydberg atoms for kHz electric field sensing, specifically focusing on the ULF, VLF, and LF bands. Our research demonstrates that the use of built-in parallel electrode plates can effectively overcome shielding effects, providing a novel technical approach for electric field measurement in these frequency ranges. By enhanced with a DC electric field, we have done measurements of 1kHz, 10kHz and 100kHz signal fields with the corresponding sensitivity of 5.7μV/cm/√Hz, 2.2μV/cm/√Hz and 0.95μV/cm/√Hz. To our knowledge, this is the first publication on the sensitivity of Rydberg-atom sensors in the VLF and LF bands. Our results represent that kHz field measurement based on Rydberg-atom sensor has exceeded the limit of dipole antennas with the same size. Rydberg sensing technology is capable of leveraging ULF, VLF, and LF fields for applications in remote timing, auxiliary navigation, as well as radio communication.

**Disclosures.** The authors declare no conflicts of interest.

**Data availability.** Data underlying the results presented in this paper are not publicly available at this time but may be obtained from the authors upon reasonable request.